# Radiofrequency and Mechanical Tests of Silver Coated CuCrZr Contacts for the ITER Ion Cyclotron Antenna

J.Hillairet[1], Z.Chen[1], G.Lombard[1], J.M.Delaplanche[1], K.Vulliez[2], Q.Yang[5], B.Beaumont[3], F.Calarco[3], N.Charabot[1], F.Kazarian[3], P.Lamalle[3], J.M.Bernard[1], V.Bruno[1], J.C.Hatchressian[1], R.Laloo[4], P.Mollard[1], Y.Song[5], V.Turq[4], R.Volpe[1],

[1] CEA, IRFM, F-13108 Saint-Paul-Lez-Durance, France

[2] Laboratoire d'étanchéité, DEN/DE2D/SEAD, CEA, 2 rue James Watt 26700 Pierrelatte, France

[3] ITER Organization, Route de Vinon-sur-Verdon, CS 90 046, 13067 St. Paul Lez Durance Cedex, France

[4] Institut Carnot CIRIMAT, UMR CNRS-UPS-INP 5085, Université Paul-Sabatier, 118 route de Narbonne, 31062 Toulouse cedex 9, France

[5] Institute of Plasma Physics, CAS, Hefei, Anhui 230031, China

*Abstract*

The ITER Ion Cyclotron Resonance Heating (ICRH) system is designed to couple to the plasma 20 MW of RF power from two antennas in the 40-55 MHz frequency range during long pulses of up to 3600 s and under various plasma conditions with Edge Localized Modes. Radio-Frequency (RF) contacts are integrated within the ITER ICRH launcher in order to ensure the RF current continuity and ease the mechanical assembly by allowing the free thermal expansion of the Removable Vacuum Transmission Line coaxial conductors during RF operations or during 250°C baking phases. A material study has been carried out to determine which materials and associated coatings are relevant for RF contacts application in ITER. In parallel, RF tests have been performed with a new prototype of Multi-Contact® LA-CUT/0,25/0 contacts made of silver-coated CuCrZr louvers. During these tests on a RF vacuum resonator, currents between 1.2 kA and 1.3 kA peak have been reached a few tens of times in steady-state conditions without any visible damage on the louvers. A final 62 MHz pulse ending in a 300 s flat top at 1.9 kA resulted in severe damage to the contact. In addition, a test bed which performs sliding test cycles has been built in order to reproduce the wear of the contact prototype after 30 000 sliding cycles on a 3 mm stroke at 175°C under vacuum. The silver coating of the louvers is removed after approximately a hundred cycles whilst, to the contrary, damage to the CuCrZr louvers is relatively low.

## 1 Introduction

The usage of electrical contacts for ICRH in nuclear fusion research at RF frequencies has been investigated and used in various fusion laboratories. The goals of these contacts are generally to ease the assembling and/or sliding movements of the structures and to lower the thermo-mechanical stress. Depending on the purpose, these sliding contacts can be located into a pressurized or a vacuum environment. CEA or IPP internal or EFDA-funded studies have been made for antenna R&D like for the JET-EP antenna or for W7-X [1], as well as in RF accelerators [2,3]. Standard commercial products Multi-Contact LA-CUT are for instance used inside the JET "ITER-like" antenna between the capacitors and the straps [4].

The ITER Ion Cyclotron Resonance Heating and Current Drive (ICRH&CD) system is designed to couple to the plasma 20 MW (10 MW each from two antennas) of RF power in the 40-55 MHz frequency range in continuous wave operation (3600 s) [5–7]. The RF power, generated by nine 3 MW sources (four per antenna and one spare)[8], is carried though pressurized coaxial lines to 8 feeds per antenna. The ITER ICRH antenna design relies on electrical sliding contacts located in a few locations of the antenna on both inner and outer conductors [9,10]. These contacts ensure the current continuity and allow the antenna assembly by lowering thermo-mechanical stresses of the Removable Vacuum Transmission Line (RVTL) coaxial conductors during the RF operations. They are required to be used in steady-state conditions and under the machine high vacuum ($10^{-3}$ – $10^{-7}$ Pa). The temperature environment of the contacts depends on the machine regime: during RF operations, the antenna in which the contacts are located is cooled by a 70–90°C water network while during baking operations (no RF) the temperature rises to 250°C. These RF contacts have to withstand maximum peak current of 2.5 kA in steady-state at 40-55 MHz, on various diameters. In terms of current densities, defined as the ratio between the current and the circumference of the contact strip, this represents between 3.1 kA/m and 4.8 kA/m depending on their location in the antenna. The highest current density corresponds to the inner conductor connection between the 4-ports junction and the RVTL for a 2 kA peak current on a 132 mm diameter at 55 MHz. The thermal expansion during RF operation on a RVTL is estimated to 2 mm, at low speed (~0.05 mm/s). During assembly or intervention, the total stroke is about 50 mm at the remote handling system speed (few mm/s) and under atmosphere. Since no routine maintenances are scheduled on the antenna, these contacts must operate during its whole lifetime, required to be at least 20 years. Thus, the reliability of the ICRH antenna is directly linked to these contacts, which have to withstand 30 000 cycles over this period. For this reason, the contact design and materials require extensive characterisations.

CEA had setup a steady-state test bed, an actively water cooled T-resonator, which allows testing coaxial components up to 45 kV and 2.25 kA in ITER relevant conditions of vacuum and temperature [11]. In a previous test campaign, a design proposed by the CYCLE consortium using brazed lamellas supported by a spring [9], was tested in the T-resonator up to 1.7 kA during 1200 s but failed for larger current values due to a degradation of the contacts [12]. During previous tests on the T-resonator or in other testing devices, higher current densities have been reached when using commercial contacts such as Multi-Contact LA-CUT [11]. A new qualification campaign has been setup in 2015-2016 in order to test a new Multi-Contact LA-CUT prototype made of 30 μm silver coated CuCrZr louvers. This prototype has been assembled into the RF resonator in order to test it under ITER relevant conditions. The prototype has also been assembled into a static electromechanical load frame. The load frame was first used to measure the required insertion forces under atmosphere. Then, the sliding forces in an ITER relevant vacuum and 175°C environment were studied. Up to 30 000 cycles on a 3 mm stroke have been performed in vacuum and at 175°C.

This paper reports the results of this test campaign and is organized as follows. The section describes the RF contact challenging design features and the previous results achieved so far in the ICRH context. The section 2 details the selection of the best suited materials for ITER RF contact applications. The section 3 reports the results of the RF tests performed in the CEA resonator with the Multi-Contact LA-CUT prototype. Finally, the section 4 reports the mechanical endurance tests.

## 2 Contact Material Selection for ITER

### 2.1 Base materials

Two types of base materials are relevant in the context of RF contact on ITER ICRH antenna: i) the contact louvers material (bulk) and ii) the base material of the facing conductor. For the facing conductors (or the contact holder), common materials are stainless steel and copper alloys. Stainless-steel has good mechanical strength and is a common material for manufacturing but its main drawback is its high electrical and thermal resistivity. In order to increase both the electrical and thermal conductivity, CuCrZr alloy is a good structural material commonly used in fusion engineering [13,14] .

On the shelf LA-CUT louvers are made of either copper or copper-beryllium [15]. However, copper is known to seriously creep in high temperature conditions, for instance during the thousands of hours of ITER baking phases at 250°C. Since creeping would impact the contact pressure applied to the conductor and degrade the contact quality with time[16], pure copper louvers must be discarded. Copper-beryllium (CuBe), copper-nickel-beryllium (CuNiBe) and copper-chromium-zirconium (CuCrZr) alloys have lower creep rate as well as higher yield strength [17,18]. The material properties of these alloys highly depends on their manufacturers and grades, but in a general way the yield strength of CuBe (>700 MPa) is higher than CuCrZr (~450 MPa) and its thermal conductivity is lower (from 50 to ~200 vs 320-340 W/m.K) [18]. In addition, in a fusion reactor, high energy neutrons will produce Ni, Zn and smaller amount of Cobalt in copper and copper alloys, producing a significant decrease of the electrical and thermal conductivity [18–20]. At the contrary of CuCrZr, CuBe and CuNiBe contains Nickel (between 1.4 to 2.2%) and Cobalt (between 0.30 to 0.60%), which may induce higher intensity gamma-ray emitter [60]Co after neutron irradiation. Depending of the total amount of these products, CuCrZr may be preferred to CuBe/CuNiBe in this context. Moreover, because of low fracture toughness and tensile elongations for temperatures higher than 250°C, CuNiBe appears to be not a suitable option unless improved thermomechanical treatment conditions are found [21]. For the latter reasons, the choice of a CuCrZr alloy for the contact louvers material has been made.

### 2.2 Coating Materials

Early current connector designs implementing metal-metal contacts were plagued with high friction, wear and heat generation mainly attributed to electrochemical oxidation and arcing [28]. For the electrical contacts working at high temperature and high vacuum, wear performance becomes critical and coatings can be used to improve their electrical performance and life-time. For RF applications, dozens of micrometers thicknesses are enough for contact materials, which is much lower than their required structural dimensions. Moreover, the geometry of the RF contact is usually complex and good uniformity of coating is required for the whole RF contact component. For these reasons, electroplating is generally used for RF contact coating.

The ideal vacuum electrical contact coating materials should have the following characterizations: high electrical conductivity, high thermal conductivity and good structural stability under high temperature, low friction coefficient and high wear resistance even under high temperature and high vacuum and if possible economic. It is difficult to find a material to meet all the above requirements, which are sometimes antagonistic. Compared with single electrodeposited metals, alloy deposition often provides superior properties. With certain composition ranges, the alloy electroplating layers can be harder, denser and have higher resistance to corrosion and wear.

Silver is a well-used contact material, since it has the highest electrical and thermal conductivity of all metals (resistivity: 1.67 µΩcm) and is also resistant against oxidation. Electroplated silver is a worthy coating for temperatures less than 160°C with good fretting performance and excellent electrical properties [23,24]. At temperatures higher than 100°C, recrystallization at temperatures decreases the hardness to lower than 90 HV 10, which can induce wear, cold welding and material migration. Coating thicknesses up to few tens of micrometers are generally deposited on the electrical contacts designed to withstand very high normal force (10-100 N), which is the case in this application. For these reasons, a thick silver-coating (>10µm) has been selected as a functional coating of the RF contact louvers.

## 3   RF Tests

### 3.1   RF Resonator Setup

In the framework of the design of the IC H&CD antenna, RF tests have been made on a 30 µm silver-coated Multi-Contact LA-CUT prototype with CuCrZr louvers. The RF resonator test bed setup referred in this paper is introduced in references [11,12]. The setup, illustrated in Figure 1, consists in a rigid coaxial resonator, made of two main branches approximatively 2.2 meters long. Both branches are RF powered from a RF source feeding a T-junction and are ended by adjustable short-circuits referred as *trolleys* hereafter. Such sliding short-circuits allow tuning the resonator matching configuration, in addition to the possibility to tune the RF input power frequency. A vacuum feedthrough is located upstream of the T-junction, which allows the T-resonator to be vacuum pumped. Electric heating cables were inserted into both branches outer conductors for baking purposes.

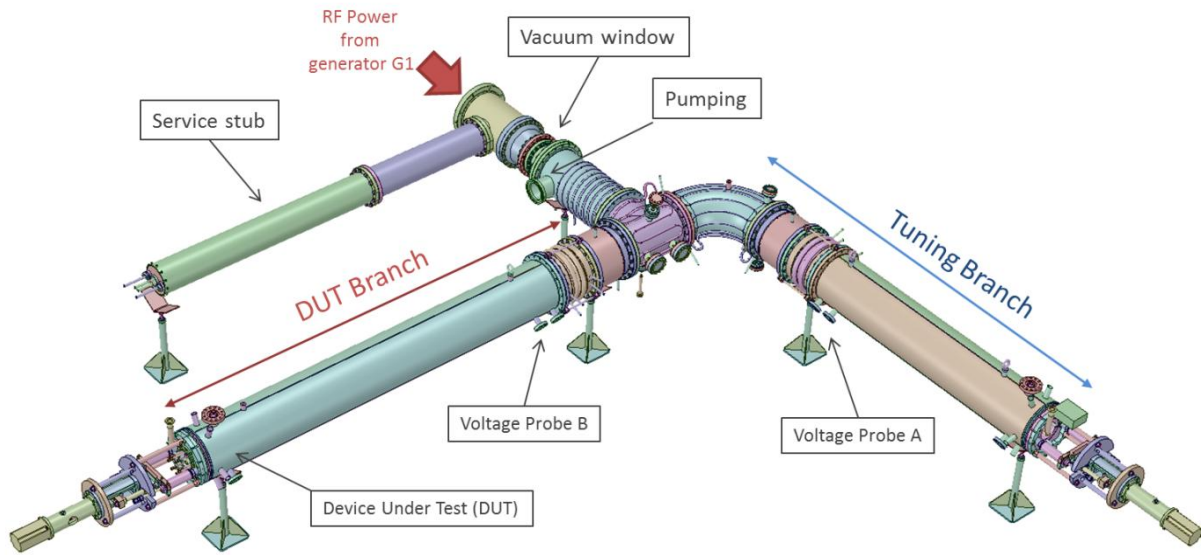

**Figure 1. General description of the RF resonator**

All the inner and outer conductors of the resonator are actively water cooled by a cold water loop (4 bar/20°C). The resonator, located in the TITAN facility [25], can also be connected to a hot water loop compatible with ITER specifications (up to 44 bar/250°C). The water cooling temperature during RF operations in the ITER IC antenna is estimated to be 90°C at the contact location. In order to reproduce the ITER conditions, the contacts under test have been assembled on a copper coated stainless-steel holder, ending the inner conductor of the DUT branch. Both this inner conductor extension and the trolley have been connected to the hot water cooling loop, tuned to a 90°C inlet temperature. The Device-Under-Test (DUT) branch is optimized to maximize the current at the short-circuit. The contact under test connects the inner conductor to the trolley, where the current density is the maximum. The connexion from the trolley to the outer conductor is made with standard (silver coated copper louvers) LA-CUT. The inner conductor diameter at the contact under test is 128 mm while the outer conductor diameter is 219 mm (Figure 2). The current density at the outer conductor is thus approximatively half of the current density at the inner conductor (128/219=0.58). A similar setup is used on the opposite branch, with a reduced current at the short circuit due to the branch characteristic impedance configuration.

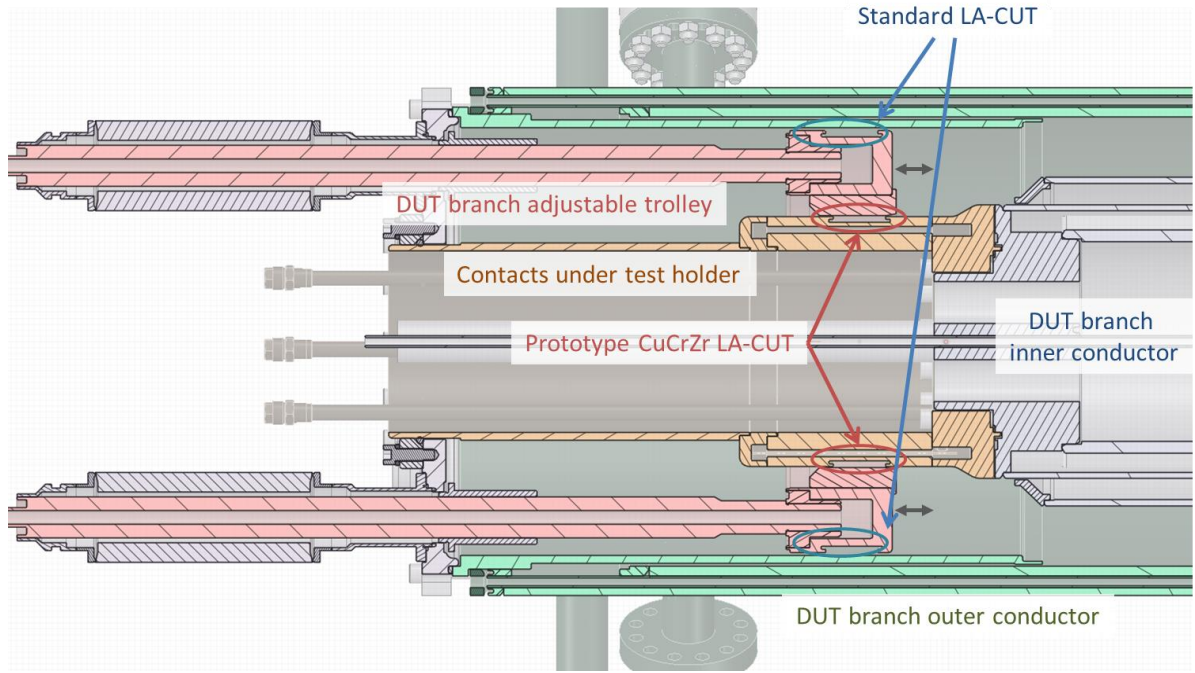

**Figure 2. Geometrical Details of the Device Under Test short-circuit equipped with prototype Multi-Contact CuCrZr LA-CUT.**

## 3.2 RF Resonator Modelling

In order to match the resonator, both short circuit lengths can be tuned as well as the generator RF frequency, which allows a large operational flexibility. Voltage is measured in two locations of the resonator (cf. Figure 1) using voltage probes similar to the ones used in the JET and Tore Supra ICRH antennas [26]. The current at the short circuits is deduced from the measured voltages using a RF model of the resonator. To do this, both a lump [27] and a full-wave model in ANSYS HFSS and Designer have been created. These models parameters are the two short-circuit distances and the two equivalent resistances presented by each short circuits. Moreover the material conductivities given in literature tables are given for bulk materials and are generally optimistic when dealing with surface coatings. Thus, the Ohmic losses in the resonator model need also to be slightly increased to match reality, by a coefficient to be determined. In order to deduce all these quantities, first a low power measurement is performed on the matched resonator. The lump model short resistances and propagation losses are then optimized in order to match the measurements. The result of the optimization is illustrated in Figure 3 where the transmission line model fits perfectly the measurements for equivalent resistances of 3.4 and 6.8 mΩ for the DUT and CEA short circuits respectively and additional propagation losses of +4.6%. In this case, the resonator has been matched around 62 MHz. The choice of a frequency higher than the ITER bandwidth aims to maximize the RF losses and thus providing an additional safety factor.

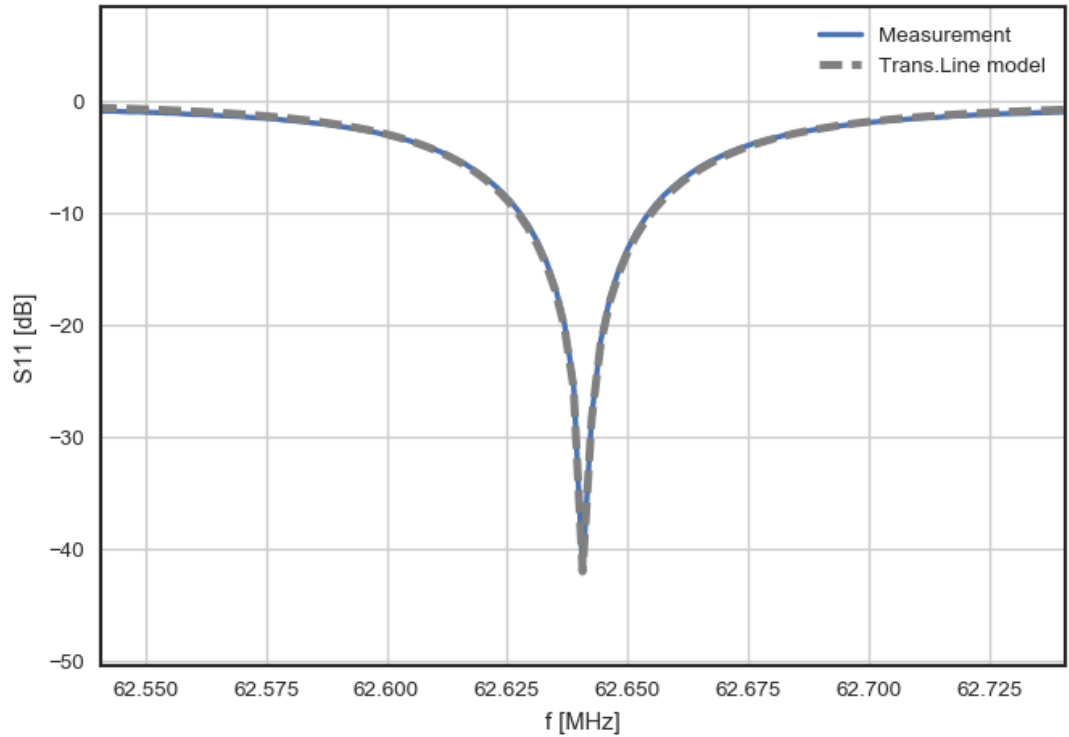

**Figure 3. Resonator input voltage reflection coefficient ($S_{11}$) when matched at 62.64 MHz. The transmission line model of the resonator is fitted to the measurements in order to deduce the equivalent resistances of both shorts.**

Once the short circuit resistances are known, the model can determine the voltage and the current everywhere in the resonator for a given input power and short-circuit lengths. Figure 4 illustrates the voltage and current distribution for an input RF power of 60 kW. A maximum voltage of 40 kV is reached in the CEA branch while a maximum current of 2 kA is reached at the DUT short circuit.

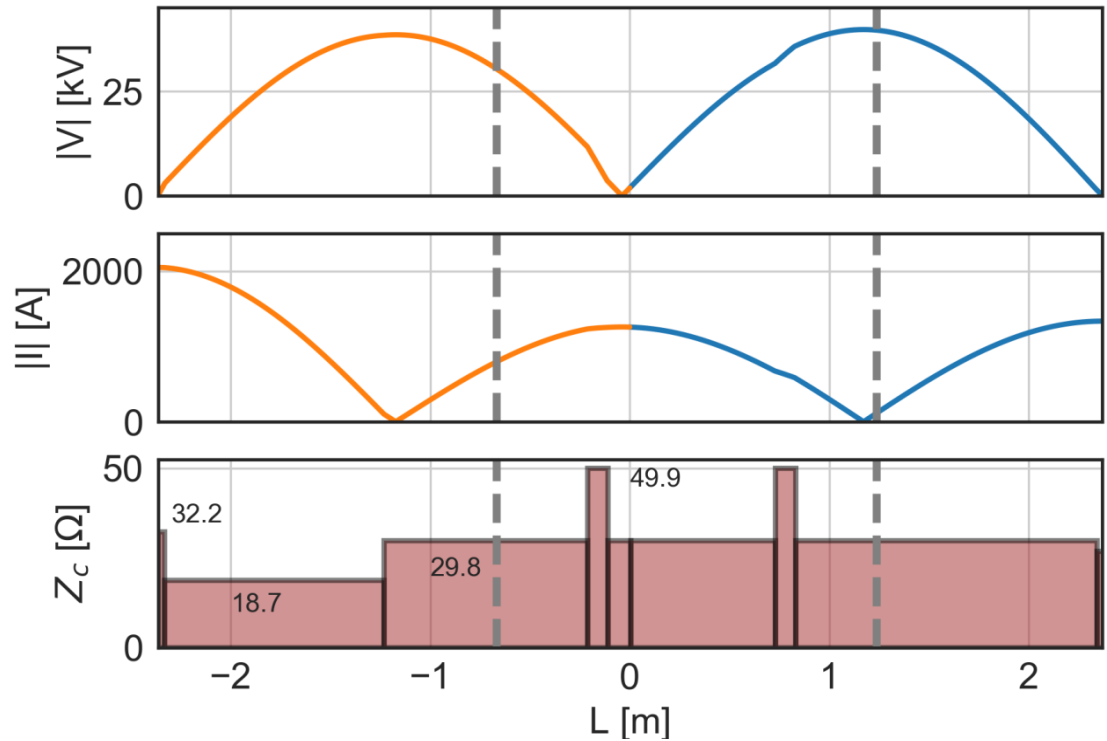

**Figure 4.** Voltage and currents inside the resonator for an input power of 60 kW at 62.64 MHz. L=0 indicates the location of the T-junction. Orange lines correspond to the DUT branch (L<0) and blue lines to the tuning (CEA) branch (L>0). Grey dashed vertical lines illustrate the locations of the voltage probes. The resonator characteristic impedances are illustrated in the third graph.

The Figure 5 illustrates the current density calculated from the ANSYS HFSS full-wave modelling of the resonator. The current density is defined as the RF peak current divided by the circumference of the surface. In this model, the resonator is modelled with a coaxial port at each end. Both short circuits are independently modelled in full-wave for a various set of lengths and frequencies. The circuit connexion and the optimization of the whole system are performed in ANSYS Designer, which allows to solve each short-circuit independently and thus reduce the computation effort. With this full-wave approach, one can estimate the heat losses on the short circuit geometry, especially on the contact louvers, a quantity which is not possible to measure.

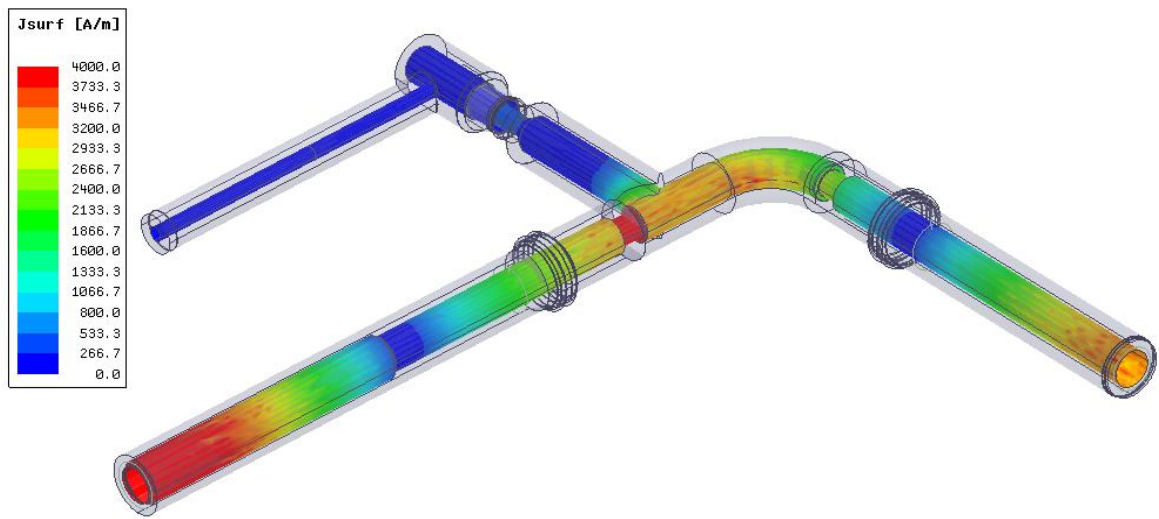

**Figure 5. Current density in A/m on the inner conductor in the full-wave model of the matched resonator excited by 60 kW at 62.64 MHz. Short-circuits at the end of both branches are not illustrated.**

## 3.3 RF Test Results

Prior to RF tests, the resonator has been pumped and baked up to 120°C during 20 days. After baking, the pressure was $8 \times 10^{-6}$ Pa. During RF tests, the resonator was water cooled at 20-25°C while the DUT was water cooled at 90°C to mimic ITER operations. RF conditioning shots consisting in kilowatt range of power pulses were performed for a set of various frequencies around the match frequency, in order to sweep the voltage and current node locations in the resonator. Once conditioning pulses do not reflect power at the match frequency, the input power has been progressively increased during short durations. Within 90 shots, the target current specification of 2.25 kA has been reached during 100 ms, then up to 5 s by increasing duration steps. In order to increase the pulse duration, the current has been reduced to 1.2kA. When increasing the pulse durations up to 300 s, the pressure increased generally after 40 to 60 s and the shots were stopped for safety. Outgassing was time correlated with light emission, as monitored through a window by a webcam viewing the rear of the DUT contacts holder and trolley. The resonator's DUT side rear flange has been disassembled for visual inspection, but as seen on Figure 6, no degradation of the louvers was observed. The resonator was then re-assembled and pumped in order to continue RF tests.

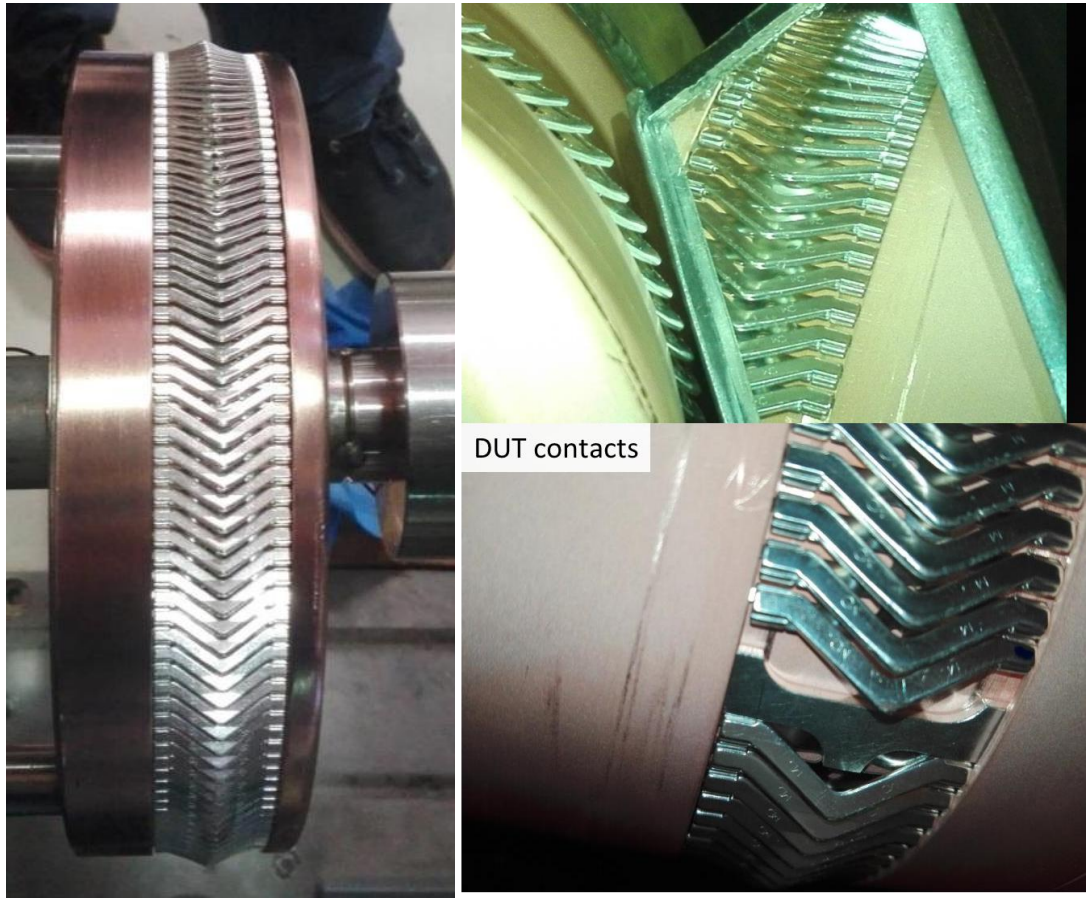

**Figure 6. Resonator Inspection at DUT side. Left: Contacts located on the DUT trolley (outer conductor). Right: CuCrZr ring (inner conductor).**

After reconditioning the resonator, the current was progressively increased in both amplitude and shot duration. During RF shot, the pressure was monitored for outgassing, but this time we allowed larger increases of the pressure, keeping an automatic safety interlock ($2\times10^{-2}$ Pa). It was found that the pressure increased after few tens of seconds, sometimes up to $10^{-2}$ Pa, but then stabilized and then decreased during the RF shot. Thus, the current was increased to 1.2kA during up to 1200 s without pressure increase larger than the safety interlock (Figure 7). Light emission was continuously monitored during these shots, starting after 60 to 120 s depending on the shot. Moreover, it was observed that the light emission changed both in brightness (increase or decrease) and locations during the shot duration as illustrated in Figure 8.

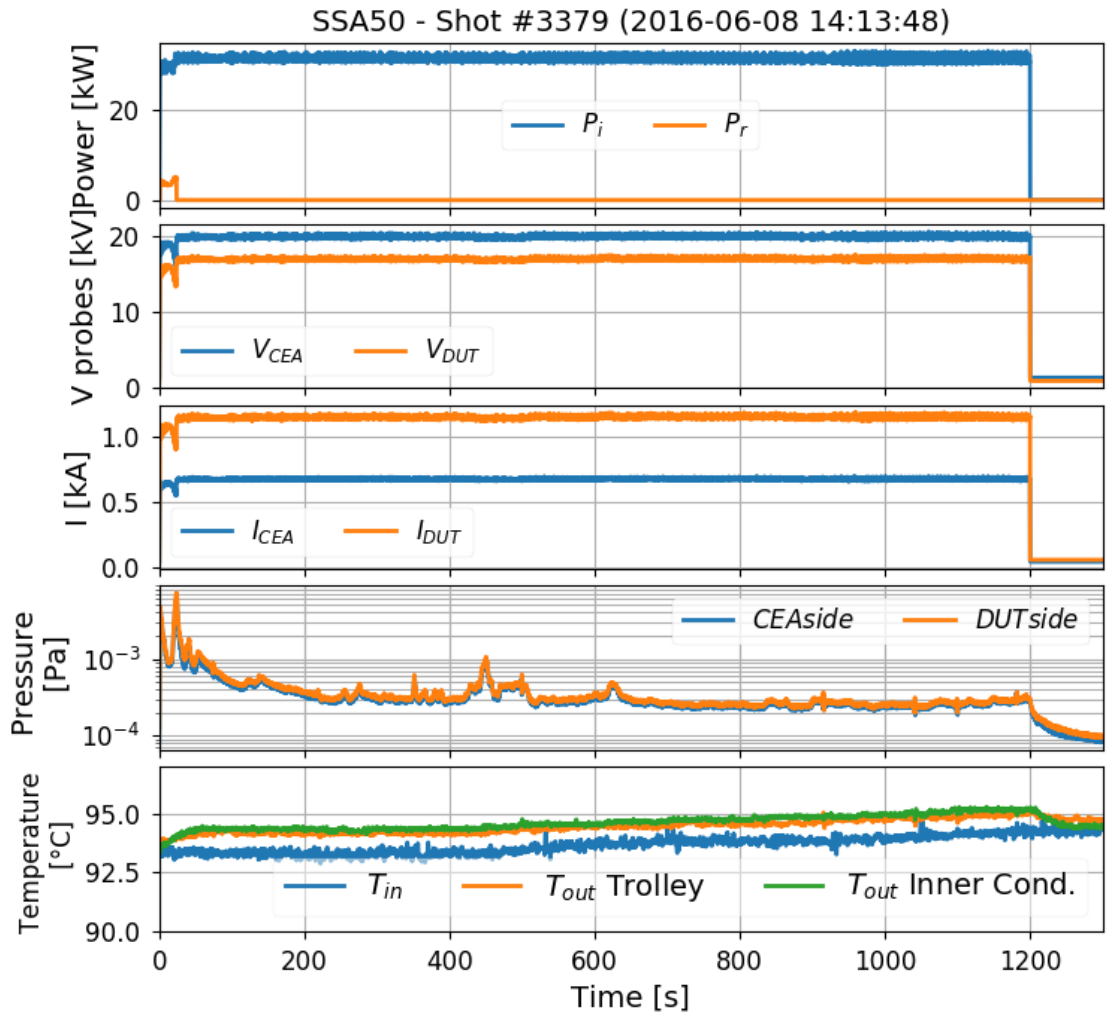

**Figure 7.** RF Resonator Shot #3379, for which 1.2 kA has been applied on the tested contact during 1200 s. From top to bottom: input and reflected power in kW, voltage measured at probes in kV, current deduced at short-circuits in kA, pressure inside the resonator and water temperature inlet and outlets.

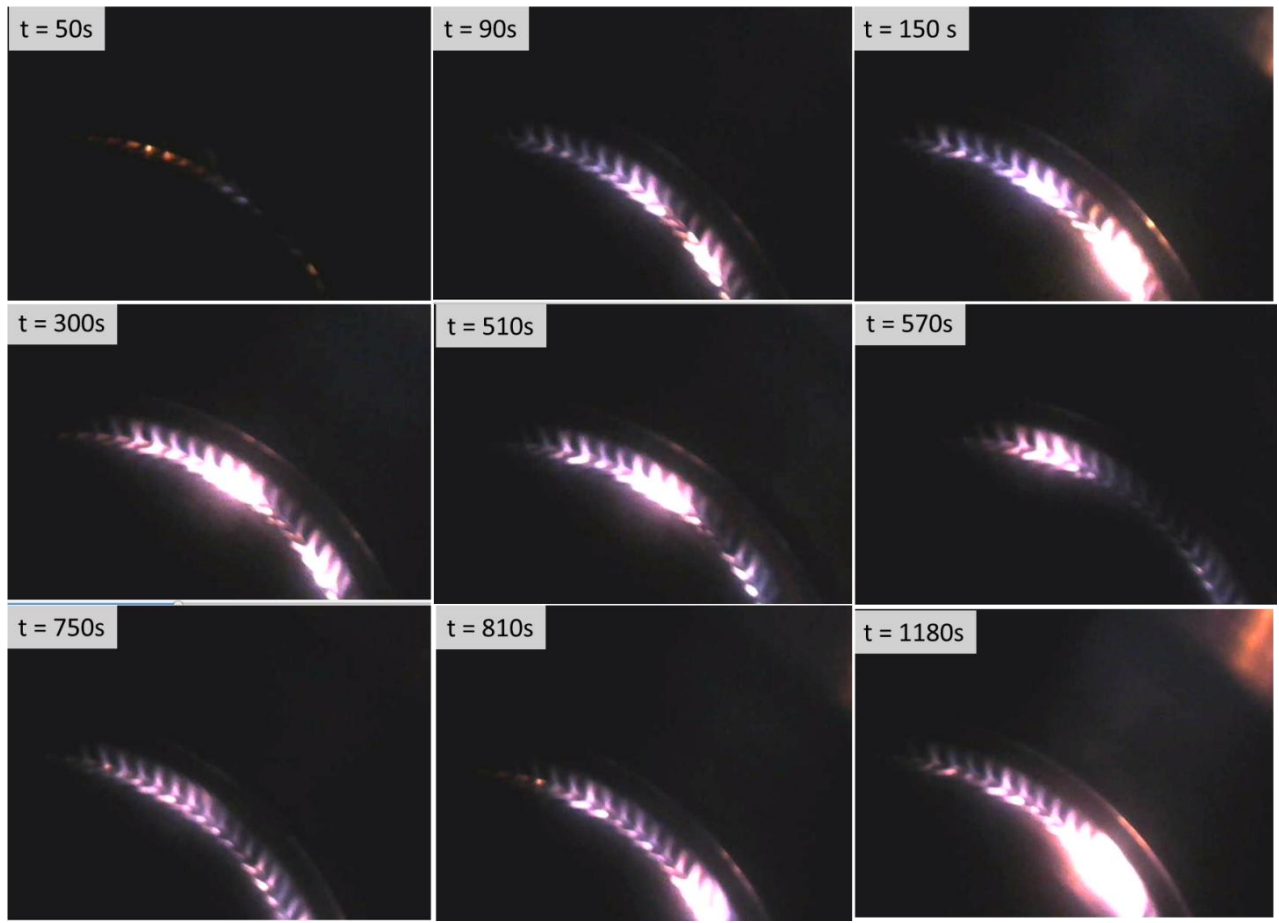

**Figure 8. Webcam snapshots during shot #3379 (1.2kA/1200s).**

At this point and for higher RF power, it was observed that the number of arc events detected (from reflected power) increased, sometimes led to the complete un-matching of the resonator, often at the beginning of the pulse but even after hundreds of seconds of RF power.

In order to reduce these effects, the input power was then ramped up instead of applying the full power from the beginning of the RF pulse. Using this strategy, the current has been increased up to 1.9 kA during 300 s (Figure 9). After this last pulse, outgassing increased up to $8 \times 10^{-2}$ Pa during the cool down of the DUT and the resonator was detuned. Outgassing was severe for the following shots and it was not possible to continue RF tests.

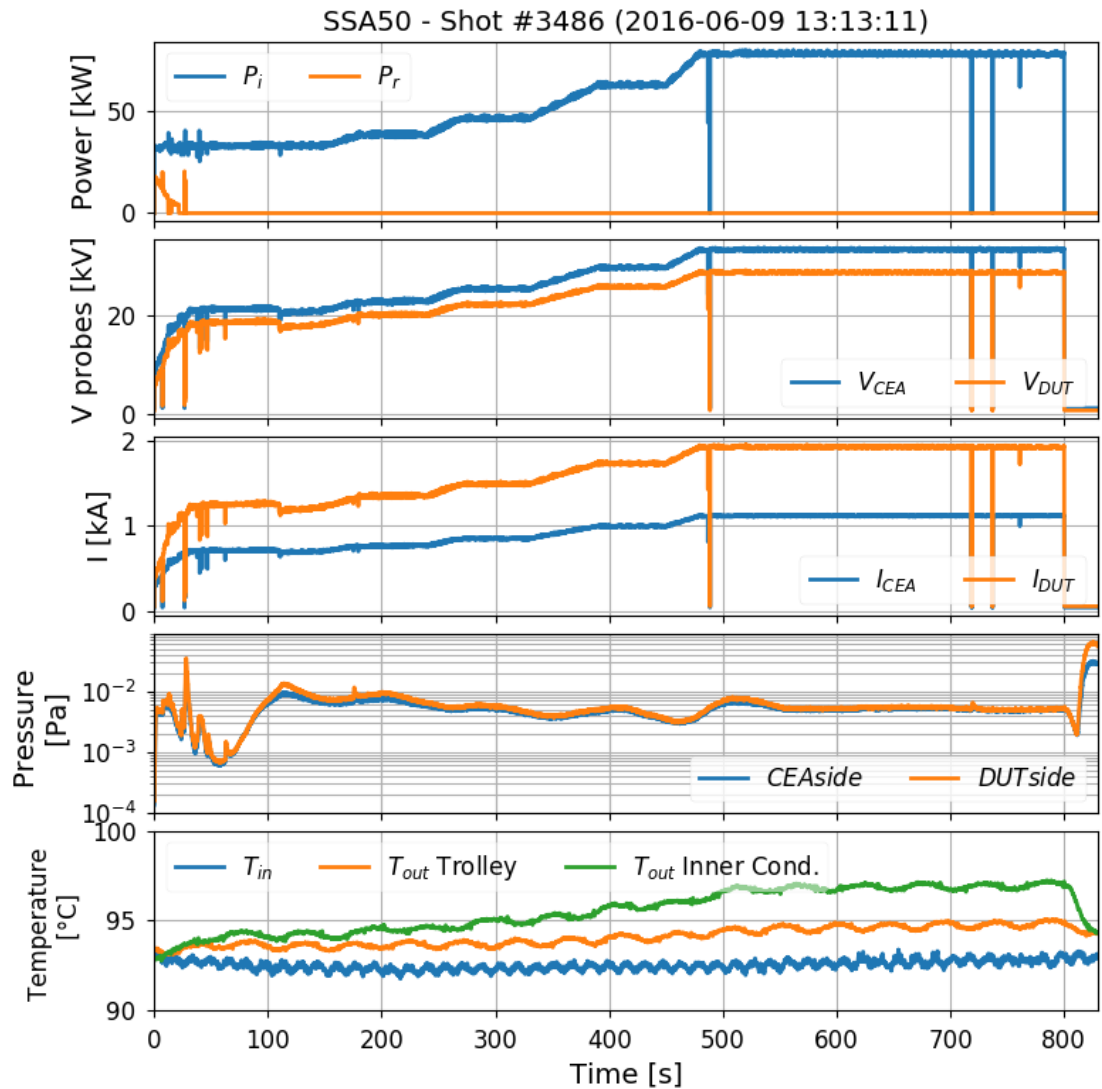

**Figure 9.** RF Resonator Shot #3486, for which a current of 1900 A has been applied on the tested contact during 300 s. The subfigure content is the same as Figure 7.

The inspection of the DUT side of the resonator showed that the DUT contact louvers were burned and melted in various places, and burn/arc traces were also observed on the trolley contact louvers.

The Figure 10 illustrates the maximum current achieved on the contact under test versus the flat-top current duration in seconds. From this figure, few elements can be highlighted. Firstly, the highest mean pressure shots come from either RF conditioning or high current/long duration shots. Secondly, the highest current shots with low mean pressure have been performed during short durations (1 to 10 s), at the exception of the repeated 1.2-1.3 kA shots. Finally, increasing shot duration is correlated with the increase of the mean pressure. However, there is a bias in this deduction, since the resonator has been open to air for inspection after the first week of test and which degraded the vacuum conditioning.

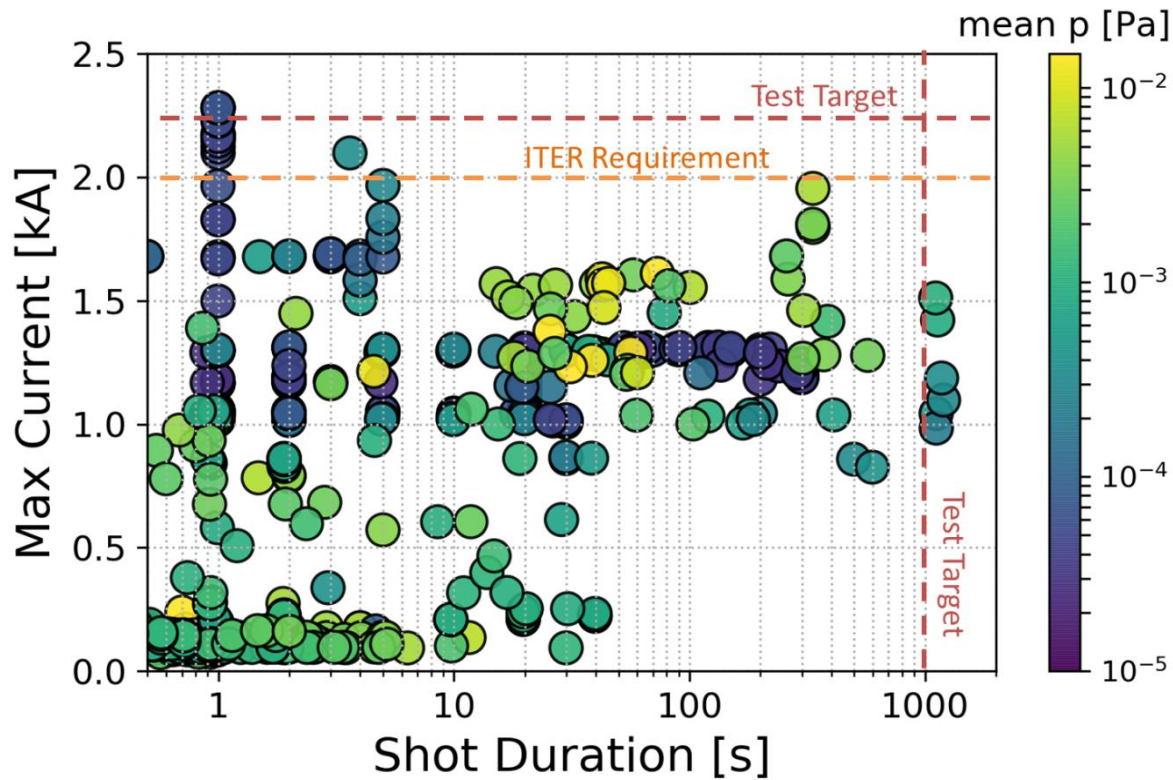

**Figure 10. Maximum current achieved during the flat-top phase on the contact under test versus the total shot duration in seconds (in log scale). The marker colour corresponds to the average pressure during (and few seconds after) the shot or to the shot duration. The RF test ultimate target corresponds to the intersection of the red dashed lines.**

The target performance, i.e. a peak current of 2.25 kA at 62 MHz, has been reached only on short pulses duration (below 10 s). Currents between 1.2 kA and 1.3 kA have been reached in steady-state conditions (with duration higher than 60 s up to 1200 s) without any problem monitored on RF or pressure sensors. An inspection of the contacts has been performed after having reached currents lower than 1.3 kA and do not show any visual damage. Dozens of shots have been performed at currents between 1.4 kA - 1.7 kA in steady-state regime (higher than 60 s and up to 1200 s) and difficulties to ramp-up the current were often found due to arcs. The best strategy found to reach these values was to progressively increase the RF power. Thermal effects seem to affect the way the RF current propagates (path and amplitude), since different light emission regimes were observed during same long duration shots. To continue on thermal effects, it was found that for current between 1.4 kA and 1.7 kA, less abnormal events (such as arcs) were found after ~60 to 100 s of RF pulses. Except for few cases, once the shot duration exceeded 120 s, they were generally not manually aborted.

The RF current has been increased up to a maximum of 1.9 kA during 300 s, but it was not possible to increase the current to larger values. The targeted steady-state RF tests specifications on the contact have not been achieved, due to the failure of the contacts at lower current values (1.9 kA). The second inspection performed at the end of the second week of RF tests, shows that both the CuCrZr louvers and the steel part of the contact band melted in some locations (Figure 11). We conclude that damages other than normal ageing happened for currents larger than 1.3 kA.

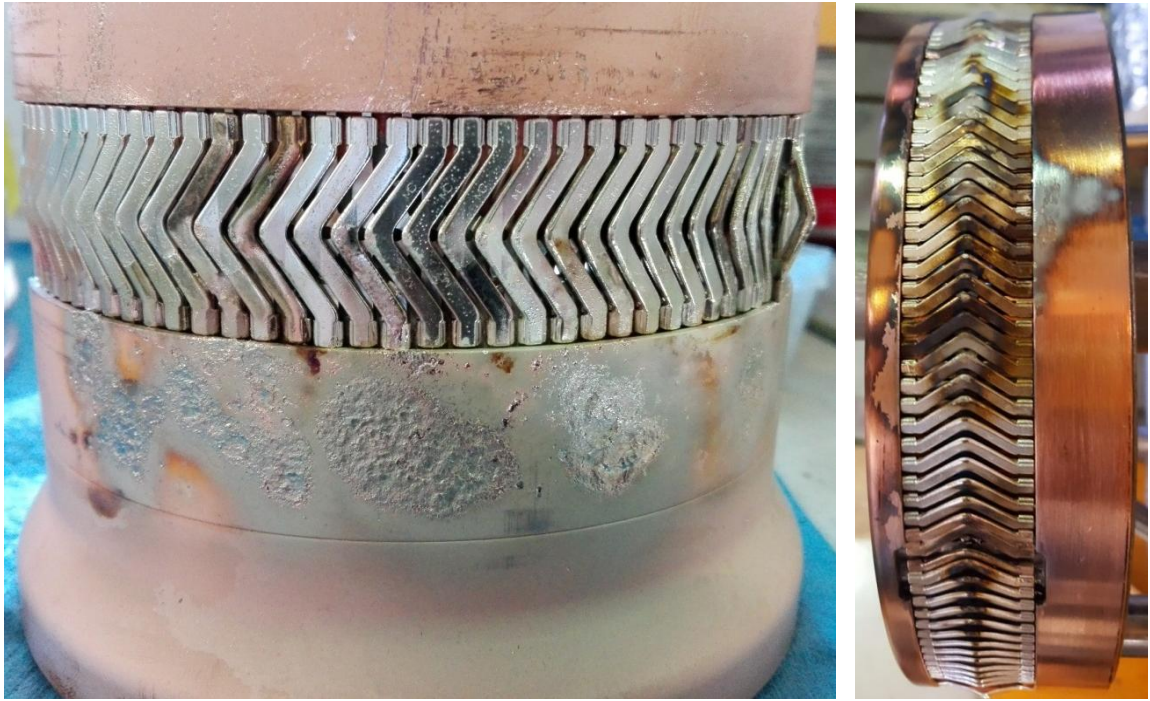

**Figure 11. Pictures of the RF contacts after failure. Left picture: inner conductor DUT contact holder. Right picture: outer conductor contacts on sliding trolley.**

# 4 Mechanical Tests

In order to measure the sliding forces, the wear performances and life-time of the coating in ITER-relevant environments, a 30 µm silver-coated CuCrZr LA-CUT prototype was assembled into a MTS© static electromechanical load frame (Figure 12). The load frame can be moved through an oven that can be baked to the required temperature, which is controlled with a precision of +/- 0.5°C. The force cell located on the top of the load frame is cooled by water and forced air to allow a stable temperature during the test. The sliding contacts (95 louvers) are inserted in a steel holder (the male part) which can be inserted in a CuCrZr ring (the female counterpart). A diaphragm bellows is assembled between both parts and the assembly can be vacuum-pumped via a turbo-molecular pump. Centring brass elements were added to the setup in order to improve the coaxial alignment between male and female parts. The position accuracy of the male part is measured within +/- 0.5% and the load is measured with a 5 kN force cell. The load frame was first used to measure the required insertion forces under atmosphere (Figure 13), then the sliding forces in an ITER relevant vacuum and 175°C environment, which corresponds to an estimate of the operational temperature of the contacts. Up to 30 000 cycles with a 3 mm stroke were performed in vacuum at 175°C.

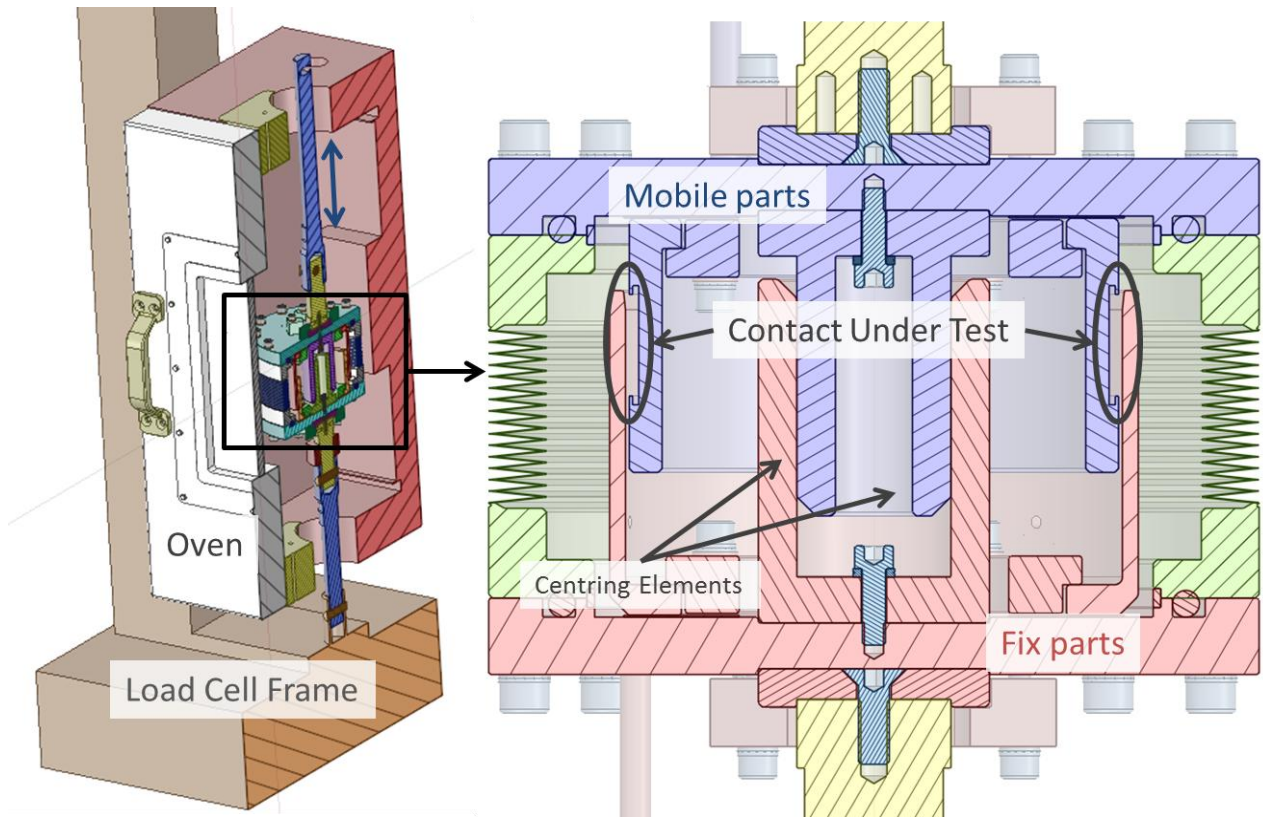

**Figure 12. CAD Illustration of the sliding LA-CUT test setup.**

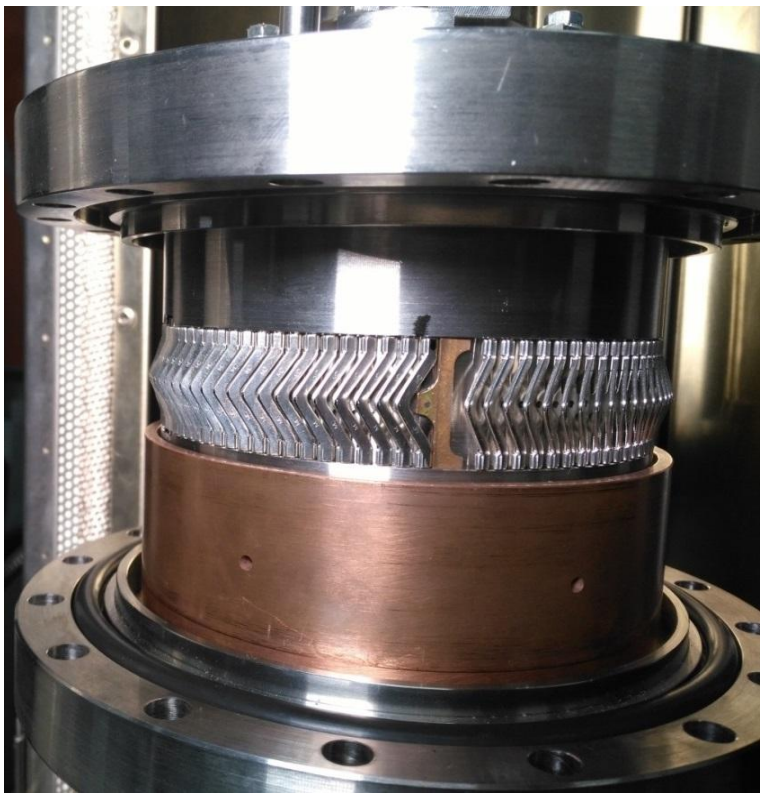

**Figure 13. Close-up picture of the setup during air insertion tests.**

The Figure 14 can be interpreted as the following: first, as the louvers start touching the CuCrZr ring, the louvers are constrained and generate a reaction force. Once the louvers are compressed, we can distinguish the sliding phases in the CuCrZr ring chamfer (red section) and in the cylindrical part (green section). It can be seen that the load trace is almost the same from the 2$^{nd}$ insertion. This is explained by an initial setup of the contact band in its groove. The required insertion force up to approximatively 700 N for the first insertion and about 550 N the following insertions, which corresponds to 5.8 N per louver. When the contact is fully engaged, the sliding force is about 200 N (2.1 N per louver). After 10 cycles, a thin layer of silver is transferred from the louvers to the CuCrZr ring.

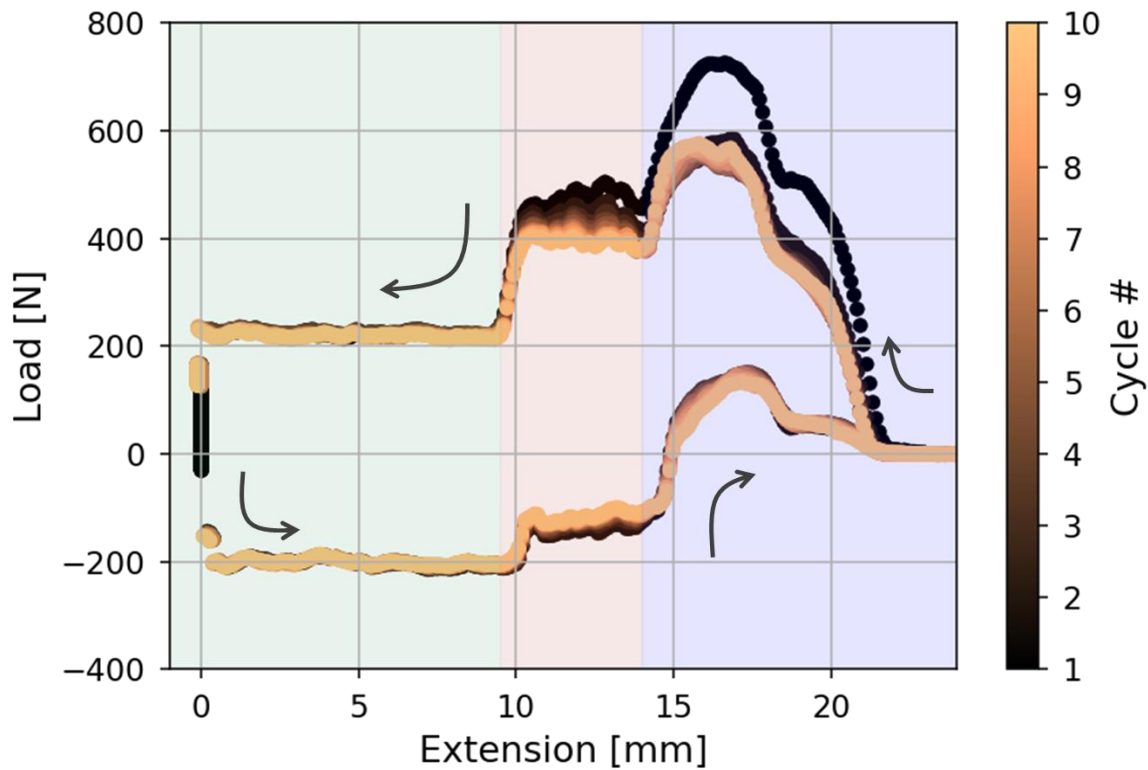

**Figure 14. Load-extension curves for the 10 insertions/extractions cycles in air. Contact insertion corresponds to upper curves while extraction corresponds to lower curves.**

The test bed was pumped and baked to 175°C with a brand new Multi-Contact band with 95 louvers assembled in the test bed. The measured pressure at 175°C is $3.2 \times 10^{-4}$ Pa. 5 000 cycles were made on an extension of 3 mm (1.5 mm back and 1.5 mm forth). The Figure 15 illustrates the load vs the cumulated distance travelled by the contact during these 5 000 cycles. It can be seen that the load is increasing then sharply decreasing over the cumulated distance of one meter, corresponding to the first 200-300 cycles. After one meter (>300 cycles), the load is almost constant and on the order of 250 N.

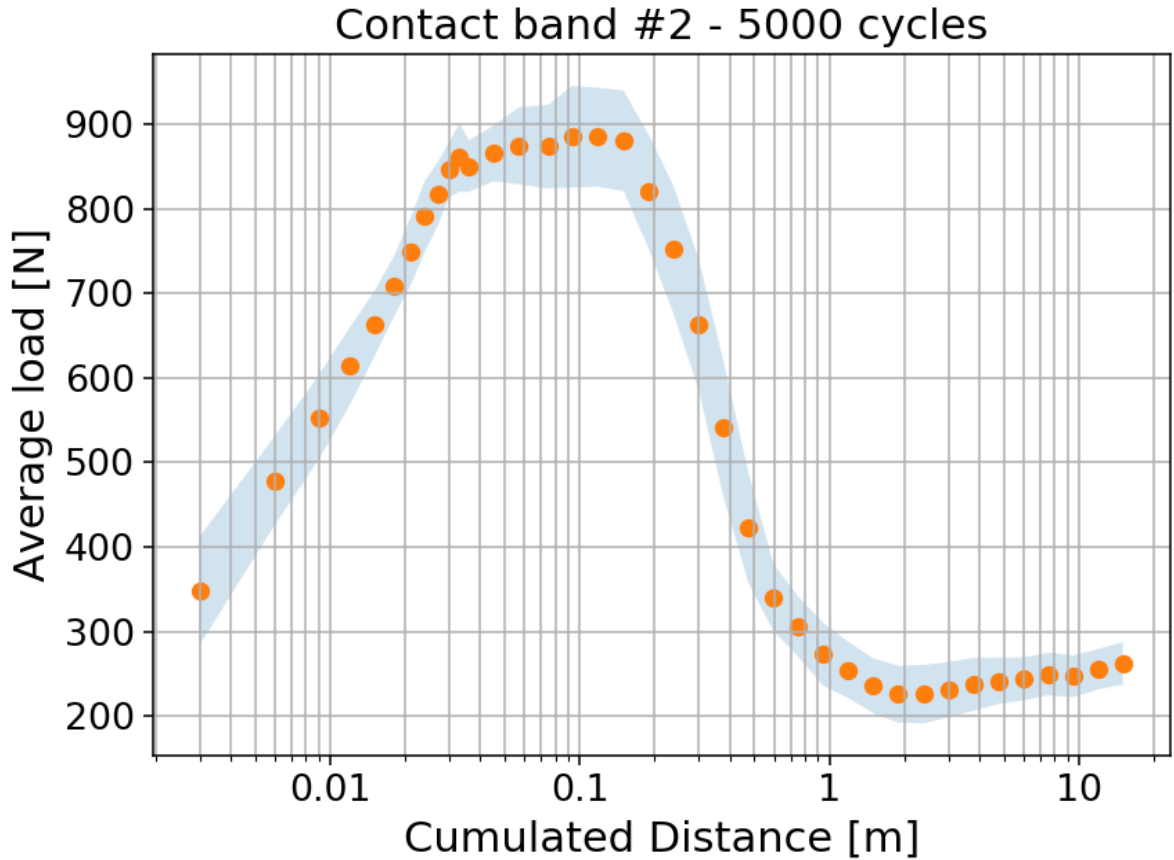

**Figure 15. Average Sliding Load versus the equivalent distance travelled by contact band.**

After 5 000 cycles, the test bed was opened and inspected. The silver coating of the contact is totally removed at the tip of the louvers. Clear scratches are visible on the CuCrZr ring, and a detailed observation of the contact shows the removal of the silver layer on the tip of louver. The area of contact (1800x900µm) after 5 000 cycles is flattened and the CuCrZr base is clearly visible. From the curves of the Figure 15, three sliding regimes can be identified. During the first tens of cycles, the sliding force magnitude increases. This increase is inferred to the machining of the silver (on the louvers) and the CuCrZr (on the ring), where the softer silver forms an interfacial layer ('buttering'). Once the silver coating fully removed in the sliding track, the abrasion of the louver tip leads to a smaller compression (by consequence the LA-CUT stainless-steel elastic skeleton is less constrained) which tends to decrease the contact force. At a certain point, an established regime is achieved (CuCrZr louvers vs CuCrZr ring).

Subsequent series of 10 000 and 15 000 cycles were made with the same contact band up to a total of 30 000 cycles (90 meters), corresponding to the required operational life of the ICRH antenna in ITER. For each series, the mechanical position of the contact stabilized in a few cycles and then the average load remained almost constant to 250 and 300 N during the remainder of the cycles. The wear on the louvers after 30 000 cycles is of the order of 150 µm (Figure 16, left picture). The wear of the CuCrZr ring after 30 000 cycles (90 meters) is of the order of 60-80 µm on a width of 0.1 mm (Figure 16, right picture). The remaining question of the impact of such scratches on the RF current flow has not been investigated and could be investigated by performing RF tests with the damaged contacts.

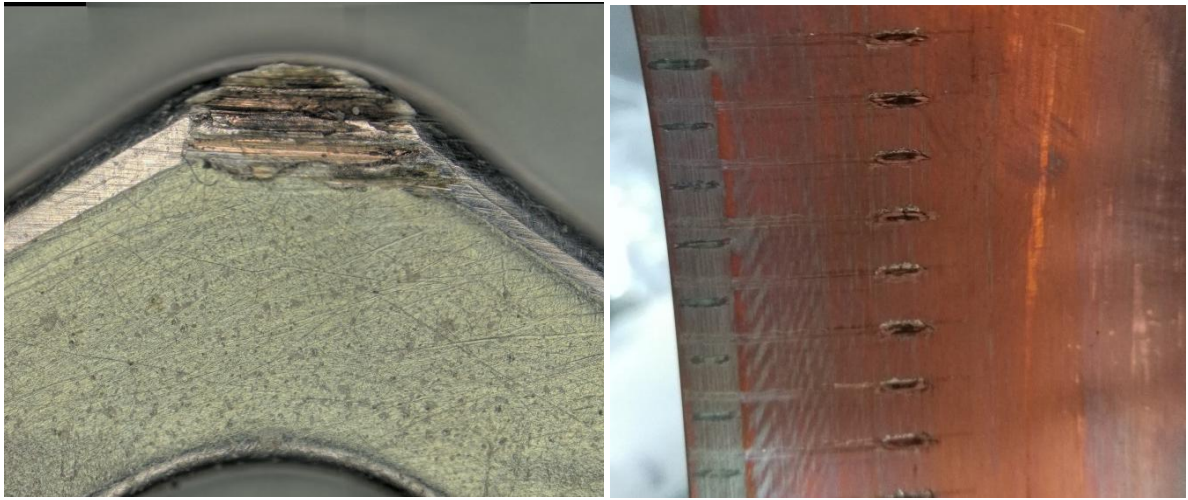

**Figure 16. Left: zoom on the tip of louver after 30 000 cycles. The wearing of the louver is of the order of 150 μm. Right: Picture of the CuCrZr ring after 30 000 cycles. The 1.5 mm long scratches depth (in the middle of the picture) is of the order of 60-80 μm on a width of 0.1 mm.**

# 5   Conclusion

High heat deposition at the contacting area is inevitable due to Ohmic losses and contact resistances. Given the large current density expected on the ITER ICRH RF contacts, high electrical and thermal conductivity materials should be used for both the contact louvers and their facing parts. Compared with pure copper, CuCrZr has much better performance of mechanical strength with a lower electrical conductivity. CuCrZr material parameters are more robust under high temperature (>100°C) neutron irradiation than other high conductivity copper alloys such as CuNiBe. For this reason, CuCrZr should be preferred for louver material. A lower resistivity functional coating such as silver coating may be added on all the louver surfaces to reduce the oxidation of the louvers made of copper alloys and to reduce the heat source due to Ohmic losses.

Radio-Frequency tests have been performed at CEA/IRFM using a RF resonator equipped with Multi-Contact® LA-CUT/0.25/0 prototype, which louvers are made in CuCrZr with a 32 μm thick silver coating. From ITER IC antenna specifications, the targeted test specifications were a peak current of 2.25 kA during 20 minutes (1200 s), under vacuum pressure below $10^{-3}$ Pa and a water cooling temperature of 90°C ± 5°C at a RF frequency of 62 MHz. No forced displacements of the RF contacts were envisaged during this test. The required current performance has been reached on durations inferior to 60 s. Currents between 1.2 kA and 1.3 kA have been reached in steady-state conditions, for durations larger than 60 s and up to 1200 s, without any problem monitored on RF or pressure sensors, despite monitoring light emission. A first visual inspection of the contacts has been performed when current on the DUT contacts hadn't been higher than 1.3 kA for long durations. Difficulties to ramp-up the current were often found due to arcs. The RF current has been increased up to a maximum of 1.9 kA during 300 s, but it was not possible to increase the current to larger values. The targeted steady-state RF tests specifications on the contact have not been achieved, due to the failure of the contacts for which both the CuCrZr louvers and the steel holder of the contact band melted in various locations. We conclude that damages other than normal ageing happened for currents larger than 1.3 kA.

From RF and thermal modelling of the contact, the following failure mechanism is proposed. Because of its relatively low electrical conductivity, high RF losses are deposited at the surface of the contact steel spring element. Since the heat accumulated in this steel part is poorly evacuated in the present design, its temperature increases which reduces its stiffness. As the contact force between the louvers and the fix part decreases accordingly, the contact resistance at the louver tips increases, leading to a vicious circle increasing the heat flux on the louvers up to melt some CuCrZr louvers. As the number of louvers in contact with the fix part decrease, the current is diverted to the remaining louvers, thus increasing again the heat fluxes and finally lead to the failure of all louvers in cascade. Reducing the RF losses on the spring holder or changing the design of the RF contact are possible solutions that could be investigated. It has to be noted that in the RF resonator, since the contact under test is located at the short circuit, steady-state arcs can be maintained without being detected since they do not mismatch the electric circuit.

The main outcome of the sliding tests is the rapid wearing of the thick silver coating. The 30 µm silver-coating made on the LA-CUT louvers is removed during approximately a hundred of stroke cycles made at 175°C under vacuum, which represents a cumulated distance travelled lower than one meter. Once this layer removed, we end up with a friction between the CuCrZr of the louver and the material facing the louver, the CuCrZr of the ring in this case. The sliding force in vacuum at 175°C is between 250 and 300 N. The damages on the bulk louvers are then relatively low, both louvers and ring material being the same. The wearing of the CuCrZr louver after 30 000 cycles (90 meters) is also low and of the order of few micrometres.

Since the coating located at the tip of the louver barely survives to more than hundreds of sliding cycles, the reason of existence these coatings could be challenged in regard of the number of the sliding cycles expected in the antenna. However, one should keep in mind that since most of the louver coating is not damaged during sliding, the gain in heat losses reduction is still important. Moreover, the Ag coating protects the CuCrZr substrate from corrosion. A remaining question is the impact of the louver tips silver-coating removal and of the generated scratches on the ring surface on the RF performances. This point could be investigated by performing RF tests with damaged contacts. Finally, in order to increase the life time of the louver coating, advanced coating made of silver or gold alloys with increased hardness could be investigated.

# 6 Acknowledgments

This work was set up in collaboration with ASIPP and ITER, with funding support of ITER Organization (SSA-50 CONV-AIF-2015-4-8). The views and opinions expressed herein do not necessarily reflect those of the ASIPP and of the ITER Organization.